\documentclass[fleqn,twoside]{article}
\usepackage{espcrc2}

\usepackage{graphicx}
\usepackage[figuresright]{rotating}
\def\bold#1{\setbox0=\hbox{$#1$}%
     \kern-.025em\copy0\kern-\wd0
     \kern.05em\%\baselineskip=18ptemptcopy0\kern-\wd0
     \kern-.025em\raise.0433em\box0 }
\def\slash#1{\setbox0=\hbox{$#1$}#1\hskip-\wd0\dimen0=5pt\advance
         to\wd0{\hss\sl/\/\hss}}
\newcommand{\be}{\begin{equation}}
\newcommand{\ee}{\end{equation}}
\newcommand{\bea}{\begin{eqnarray}}
\newcommand{\eea}{\end{eqnarray}}
\newcommand{\nn}{\nonumber}

\newcommand{\ket}[1]{\left| #1 \right\rangle}
\newcommand{\spur}[1]{\not\! #1 \,}

\newcommand{\AmS}{{\protect\the\textfont2
  A\kern-.1667em\lower.5ex\hbox{M}\kern-.125emS}}

\title{Two topics for a discussion  on the $b \bar s$ and $b \bar q$ systems }

\author{P. Colangelo, F. De Fazio and R. Ferrandes\address{INFN - Sezione di Bari, 
        via Orabona 4, I-71026 Bari, Italy}%
               }
       
\begin{document}

\begin{abstract}
The analysis  of the $b \bar s$ system is an important issue in the Physics programs of the hadron colliders. We discuss two different topics: the structure of the orbitally excited states and  prediction of
the rates of a class of non leptonic $B_s$ decays.

\vspace{1pc}
\end{abstract}

\maketitle
Experiments at the hadron colliders can analyze different aspects of the heavy meson systems, both
in the strong, both in the weak sector. For such systems, predictions are possible using
symmetries and the available data. We present two examples of experimental interest:
the properties of the orbitally excited $b \bar s$ ($b \bar q$) resonances and  the rates of a class of nonleptonic $B_s$ decays.

\section{PROPERTIES OF  ORBITALLY EXCITED $b\bar s$, $b \bar q$ STATES}

A theoretical framework to describe the excited $b\bar s$, $b \bar q$ states is
the  heavy quark chiral effective theory, 
constructed  using the spin-flavour symmetry 
for hadrons comprising a single heavy quark, in the $m_Q \to \infty$ limit,
and the chiral symmetry valid in the massless limit for  light quarks
\cite{hqet_chir}.
Heavy $Q \bar q$ mesons are classified in 
doublets according to the
 angular momentum $s_\ell$ of the light degrees of freedom: $s_\ell=s_{\bar q}+ \ell$ ($s_{\bar q}$
is the light antiquark spin, $\ell$ the orbital angular
momentum of the light degrees of freedom relative to the heavy
quark) \cite{positivep}.
For $\ell=0$  the $ s_\ell^P={1 \over 2}^-$
  doublet comprises  two states with  $J^P=(0^-,1^-)$:
$P=D_{(s)} (B_{(s)})$  and  $P^{*}=D^*_{(s)} (B^*_{(s)}) $
mesons in case of charm (beauty), respectively.
 For $\ell=1$ it could be either $ s_\ell^P={1 \over 2}^+$ or $
s_\ell^P={3 \over 2}^+$.  The  two corresponding doublets have
$J^P=(0^+,1^+)$ and $J^P=(1^+,2^+)$.  We denote the 
$J^P_{s_\ell}=(0^+,1^+)_{{1/2}}$ states  as $(P^*_{0},P_{1}^\prime)$
 and  the $J^P_{s_\ell}=(1^+,2^+)_{3/2}$  states  as $(P_{1},P^*_{2})$.
The negative and positive  parity doublets are  described by the  fields
$H_a$, $S_a$ and  $T_a^{\mu}$ ($a=u,d,s$ is a light flavour index):
$H_a  = \frac{1+{\rlap{v}/}}{2}[P_{a\mu}^*\gamma^\mu-P_a\gamma_5] $, 
$S_a = \frac{1+{\rlap{v}/}}{2} \left[P_{1a}^{\prime \mu}\gamma_\mu\gamma_5-P_{0a}^*\right]$, 
$T_a^\mu =\frac{1+{\rlap{v}/}}{2} \left\{ P^{*\mu\nu}_{2a} \gamma_\nu -P_{1a\nu} \sqrt{3 \over 2} \gamma_5 \left[
g^{\mu \nu}-{1 \over 3} \gamma^\nu (\gamma^\mu-v^\mu) \right]
\right\}$,
with the various operators annihilating mesons of four-velocity $v$.
 
 \begin{table*}[htb]
    \caption{ $\lambda_{i}$ parameters,    spin-averaged masses
     and   mass splittings $\Delta_{S}$  and $\Delta_{T}$.}
    \label{lambdaHST}
    \begin{center}
    \begin{tabular}{lcccc}
      \hline 
  & $c{\bar q}$  & $c{\bar s}$ &$b{\bar q}$ &$b{\bar s}$\\ \hline $\lambda_H$ & $ (261.1 \pm 0.7\,\, {\rm MeV})^2$ &
 $ (270.8 \pm 0.8 \,\, {\rm MeV})^2$
 & $ (247 \pm 2 \,\, {\rm MeV})^2$ & $ (252 \pm 10 \,\, {\rm MeV})^2$ \\
$\lambda_S$ & $ (265 \pm 57 \,\, {\rm MeV})^2$ &$(291 \pm 2 \,\,
{\rm MeV})^2$
 &  & \\
$\lambda_T$ & $(259\pm 10 \,\, {\rm MeV})^2$ &$ (266 \pm 6\,\,
{\rm MeV})^2$ & &\\  \hline
$\overline M_{H}$ & $1974.8\pm 0.4 \, {\rm MeV}$ & $2076.1\pm 0.5 \, {\rm MeV}$ & $5313.5\pm 0.5 \, {\rm MeV}$ & $5404\pm 3 \, {\rm MeV}$ \\
$\overline M_{S}$ & $2397\pm 28 \, {\rm MeV}$ & $2424\pm 1 \, {\rm MeV}$ & & \\
$\overline M_{T}$ & $2445.1\pm 1.4 \, {\rm MeV}$ & $2558 \pm 1 \, {\rm MeV}$ & & \\ \hline
$ \Delta_{S}$ & $422\pm 28 \, {\rm MeV}$ & $348\pm 1 \, {\rm MeV}$ & & \\
$ \Delta_{T}$ & $470.3\pm 1.5 \, {\rm MeV}$ & $482\pm 1 \, {\rm MeV}$ & & \\
 \hline
    \end{tabular}
    \end{center}
\end{table*}

The octet of light pseudoscalar mesons is introduced using 
 $\displaystyle \xi=e^{i {\cal M} \over
f_\pi}$ and $\Sigma=\xi^2$; the matrix ${\cal M}$ contains
$\pi, K$ and $\eta$ fields:
\begin{equation}
{\cal M}= \left(\begin{array}{ccc}
{\pi^0\over \sqrt 2}+{\eta\over \sqrt 6} & \pi^+ & K^+\nonumber\\
\pi^- & -{\pi^0\over \sqrt 2}+{\eta \over \sqrt 6}& K^0 \nonumber\\
K^- & {\bar K}^0 &-\sqrt{\frac{2}{3}}\eta
\end{array}\right) \nonumber
\end{equation}
with $f_{\pi}=132$ MeV.
The effective QCD Lagrangian is constructed imposing
invariance under heavy quark spin-flavour transformations and chiral transformations.
The kinetic term
 \begin{eqnarray} {\cal L} &=& i\; Tr\{ {\bar H}_b v^\mu
D_{\mu ba}  H_a \}  + \frac{f_\pi^2}{8}
Tr\{\partial^\mu\Sigma\partial_\mu \Sigma^\dagger \} \nn \\ &+&
Tr\{ {\bar S}_b \;( i \; v^\mu D_{\mu ba} \; - \; \delta_{ba} \;
\Delta_S)
 S_a \}  \nn \\
&+&   Tr\{ {\bar T}_b^\mu \;( i \; v^\mu D_{\mu ba} \; - \;
\delta_{ba} \; \Delta_T)  T_{a \mu} \}   \label{L}
\end{eqnarray}
(with 
$D_{\mu ba}=-\delta_{ba}\partial_\mu+\frac{1}{2}\left(\xi^\dagger\partial_\mu
\xi
+\xi\partial_\mu \xi^\dagger\right)_{ba}$ and 
${\cal A}_{\mu ba}=\frac{i}{2}\left(\xi^\dagger\partial_\mu
\xi-\xi
\partial_\mu \xi^\dagger\right)_{ba}$) involves
 the mass splittings $\Delta_S$ and $\Delta_T$  between   positive and
negative parity doublets. They can be expressed
 in terms of the spin-averaged masses:
 $\Delta_S= \overline M_S - \overline M_H$ and  $\Delta_T= \overline M_T - \overline M_H$,
with $ {\overline M}_H= (3 M_{P^*}+M_P)/  4$, 
${\overline M}_S = (3 M_{P^\prime_1}+M_{P_0^*})/ 4$ and
${\overline M}_T = (5 M_{P^*_2}+3M_{P_1})/8$. 

At the leading order in the heavy quark expansion
the decays  $H \to H^{\prime} M$, $S\to H^{\prime} M$ and
$T\to H^{\prime} M$  ($M$ a light pseudoscalar meson) are described by
the Lagrangian terms:
\bea
{\cal L}_H&=&g \, Tr [{\bar H}_a H_b \gamma_\mu \gamma_5 {\cal
A}_{ba}^\mu ] \nn \\
{\cal L}_S&=&h \, Tr [{\bar H}_a S_b \gamma_\mu \gamma_5 {\cal
A}_{ba}^\mu ]\, + \, hc  \label{lag-hprimo}\\
{\cal L}_T&=&{h^\prime \over \Lambda_\chi}Tr[{\bar H}_a T^\mu_b
(i D_\mu {\spur {\cal A}}+i{\spur D} { \cal A}_\mu)_{ba} \gamma_5
] + hc  \nn 
\eea
where $\Lambda_\chi$ is  the chiral symmetry-breaking scale ($\Lambda_\chi = 1 \, $ GeV).
${\cal L}_S$ and ${\cal L}_T$ describe transitions of positive parity heavy mesons with
the emission of light pseudoscalars in $S$ and $D$ wave, respectively, with
coupling constants  $h$ and $h^\prime$.  

\begin{table*}[htb]
    \caption{Predicted masses of excited beauty mesons.}
    \label{bmasses}
    \begin{center}
      \begin{tabular}{ccccc}
      \hline 
 & $ B^*_{(s)0} \, (0^{+})$ & $ B^\prime_{(s)1} \, (1^{+})$ & $ B_{(s)1} \, (1^{+})$ & $B^*_{(s)2} \,(2^{+})$ \\ \hline
$b \bar q$& $5.70\pm0.025 \,{\rm GeV}$ &$5.75\pm 0.03 \,{\rm GeV}$ & $5.774\pm 0.002 \,{\rm GeV}$ & $5.790\pm 0.002 \,{\rm GeV}$ \\
$b \bar s$& $5.71\pm 0.03 \,{\rm GeV}$ &$5.77\pm 0.03 \,{\rm GeV}$ & $5.877\pm 0.003 \,{\rm GeV}$ & $5.893\pm 0.003 \,{\rm GeV}$ \\
\hline 
    \end{tabular}
    \end{center}
\end{table*}

Corrections to the heavy quark limit  induce symmetry
breaking terms  suppressed by increasing powers of
$m_Q^{-1}$ \cite{Falk:1995th}. Mass degeneracy between the members
of the  meson doublets is broken by:
\bea
&&{\cal L}_{1/m_{Q}}={1 \over 2 m_{Q}} \big\{ \lambda_H Tr [{\bar H}_{a}
\sigma^{\mu \nu} H_{a} \sigma_{\mu \nu}]  \hspace*{0.5cm}\nonumber \\
&-&\lambda_S Tr [{\bar S}_{a}
\sigma^{\mu \nu} S_{a} \sigma_{\mu \nu}] 
+\lambda_T Tr [{\bar T}^\alpha_{a}
\sigma^{\mu \nu} T^\alpha_{a} \sigma_{\mu \nu}] \big\} \nn\\
\label{mass-viol} \eea
 with  $\lambda_{H,S,T}$  related  to
the hyperfine mass splittings:
$\lambda_H =  \left( M_{P^*}^2-M_P^2
\right)/8$, $ \lambda_S = \left(
M_{P^\prime_1}^2-M_{P_0^*}^2 \right)/8$, $\lambda_T =3  \left( M_{P^*_2}^2-M_{P_1}^2 \right)/8$ .
Other two  effects due to spin
symmetry-breaking  concern the possibility that  the members of the
$s_\ell={3\over 2}^+$ doublet can also decay in S wave into the lowest lying heavy
mesons and pseudoscalars, and that a
mixing may be induced between the two $1^+$ states belonging to
the two positive parity  doublets with different $s_\ell$. The
corresponding terms in the effective Lagrangian are:
\bea
{\cal L}_{D_1}&=&{f \over 2m_Q \Lambda_\chi}Tr [{\bar H}_{a} \sigma^{\mu \nu}
T^\alpha_{b} \sigma_{\mu \nu} \gamma^\theta \gamma_5 (i D_\alpha {\cal
A}_\theta  \nn \\
&+& iD_\theta {\cal A}_\alpha)_{ba}] + hc  \,  \label{ld1} \\
{\cal L}_{mix}&=&{g_1 \over 2m_Q} Tr[{\bar S}_{a} \sigma^{\mu
\nu}T_{\mu a} \sigma_{ \nu \alpha}v^\alpha] \, + hc  \,\,\, .\label{lmix}\eea
 The mixing angle between the  $1^+$ states:
$\ket{P_1^{phys}}=\cos \theta
\ket{P_1}+ \sin{\theta} \ket{P_1^\prime}$ and
$\ket{P_1^{\prime phys}}=-\sin \theta
\ket{P_1}+ \cos{\theta} \ket{P_1^\prime}$
 can be related to the coupling constant $g_1$ and to the mass splitting:
$\displaystyle \tan \theta={\sqrt{\delta^2+\delta_g^2}-\delta \over \delta_g}$
 where
$\delta=\displaystyle{\Delta_T -\Delta_S \over 2}$ and
$\delta_g=-\displaystyle{\sqrt{2 \over 3}{g_1 \over m_Q}}$.
\begin{table*}[thb]
    \caption{Decay widths  and branching fractions of
    $J^P_{s_\ell}=(1^+,2^+)_{3 \over 2}$
     beauty mesons obtained using the theoretical masses. To compute the full  widths  we
     assume  saturation of the two-body modes.}
    \label{larghezze}
    \begin{center}
    \begin{tabular}{c c c | c c c}
      \hline 
      Mode & $\Gamma$(MeV)  & BR & Mode & $\Gamma$(MeV) & BR  \\\hline
$B_2^{*0} \to B^+ \pi^-$ & $20 \pm 5$ & $0.34$ & $B_{s2}^{*0} \to B^+K^-$ & $4 \pm 1$ & $0.37$ \\
$B_2^{*0} \to B^0 \pi^0$  & $10.0 \pm 2.3$ & $0.17$& $B_{s2}^{*0} \to B^0 K^0$  &  $4 \pm 1$ & $0.34$\\
$B_2^{*0} \to B^{*+} \pi^-$  & $18 \pm 4$ &$0.32$ & $B_{s2}^{*0} \to B^{*+} K^-$  & $ 1.7 \pm 0.4$ & $0.15$\\
$B_2^{*0} \to B^{*0} \pi^0$  & $ 9.3 \pm 2.2$ & $0.16$& $B_{s2}^{*0} \to B^{*0} K^0$  & $1.5 \pm 0.4$ &$0.13$\\
$B_2^{*0}$ & $57.3\pm 13.5$ & &$B_{s2}^{*0}$ & $11.3\pm2.6 $ &\\
  \hline
$B_1^{0} \to B^{*+} \pi^-$  & $28 \pm 6$ & $0.66$& $B_{s1}^{0} \to B^{*+} K^-$  & $ 1.9 \pm 0.5$ &$0.54$\\
$B_1^{0} \to B^{*0} \pi^0$  & $ 14.5 \pm 3.2$ &$0.34$ & $B_{s1}^{0} \to B^{*0} K^0$ & $1.6 \pm 0.4$ &$0.46$ \\
 $B_1^{0}$ & $43\pm10$ & & $B_{s1}^{0}$ & $3.5\pm1.0$ &\\
     \hline 
    \end{tabular}
    \end{center}
\end{table*}
Some of the parameters in the effective Lagrangian can be determined using
recent measurements on charmed and charmed-strange mesons
\cite{paper1}, in particular on
 two broad states  which could be identified as the
$D^*_0$ and  $D^\prime_1$ mesons  ($s_\ell^{P}={1\over 2}^{+}$)
\cite{Abe:2003zm} and on
 the two mesons $D^{*}_{sJ}(2317)$
 and $D_{sJ}(2460)$ \cite{Aubert:2003fg} which
 naturally fit in the doublet ($D^*_{s0}$, $D^\prime_{s1}$) and,  being below the  $DK$ and $D^*K$ decay thresholds, are narrow  \cite{Colangelo:2004vu}.

The two sets of measurements, together with
 the masses  reported by PDG \cite{PDG}  and
 the $B^*_s$ mass recently measured by CLEO:
 $m_{B^*_s}=5414 \pm 1 \pm 3  \ {\rm MeV}$  \cite{Bonvicini:2005ci},
 allow  to determine some  parameters  appearing in
eqs.(\ref{L}-\ref{lmix}), see
Table \ref{lambdaHST} .
For the mixing  angle between the
two $1^+$ states $D_1$ and $D_1^\prime$, considering
 the  Belle's  result  $\theta_c=-0.10 \pm 0.03 \pm 0.02\pm 0.02 \,\,rad$  in \cite{Abe:2003zm}
 and  using $\Delta_T$ and $\Delta_S$ in Table \ref{lambdaHST}
 and  $m_{c}=1.35 \,$ GeV,
we can determine  the coupling $g_1$ in (\ref{lmix}): $g_1=0.008\pm
0.006 \,$  GeV$^2$. In the beauty
system, for  $m_b=4.8 \,$  GeV,  one  obtains:  $\theta_b \simeq
-0.028 \pm 0.012\,\,\, rad$. 
 %
  \begin{table*}[htb]
\caption{$SU(3)$ amplitudes  for  $B^-$, $\overline{B}^0$ and  $\overline{B}_s^0$ decays 
to $ D_{(s)} P$, induced by $b \to c \bar d(s)$ transitions.
 $P$
is a light pseudoscalar meson.  The predicted  $\overline{B}_s^0$ branching
fractions  are also reported.}
    \label{tab1}
    \begin{center}
     \begin{tabular}{ l c c | l c c}
  \hline 
  $B^-,  \overline{B}^0$&  amplitude &  BR (exp)&
  $\overline{B}_s^0 $ &  amplitude &  BR (th)\\
  \hline
$D^0\pi^-$ & $V^*_{ud}V_{cb}\,\, (C+T)$&  $(4.98\pm0.29)\times10^{-3}$&
$D_s^+\pi^-$ & $V^*_{ud}V_{cb}\,\, T$&  $(2.9\pm0.6)\times 10^{-3}$ \\
$D^0\pi^0$ &$V^*_{ud}V_{cb}\,\, \frac{(C-E)}{\sqrt 2}$& $(2.91\pm0.28)\times10^{-4}$ &
$ D^0\overline{K}^0$ & $V^*_{ud}V_{cb}\,\,C$&  $(8.1\pm1.8)\times 10^{-4}$ \\
$ D^+\pi^-$ & $V^*_{ud}V_{cb}\,\, (T+E)$& $(2.76\pm0.25)\times10^{-3}$ & & & \\
$D^+_sK^-$ & $V^*_{ud}V_{cb}\,\,  E$& $(3.8\pm1.3)\times10^{-5}$ & & & \\
$ D^0 \eta_8$ & $ - V^*_{ud}V_{cb}\,\, \frac{(C+ E)}{\sqrt6}$&   &
$D^0 \eta_8$ &$V^*_{us}V_{cb}\,\, \frac{(2 C-E)}{\sqrt{6}}$ &  \\
$ D^0  \eta_0$ & $V^*_{ud}V_{cb}\,\,  D$&    & 
$D^0 \eta_0$ &$V^*_{us}V_{cb}\,\, D$ &  \\
$ D^0 \eta$ & & $(2.2\pm0.5)\times10^{-4}$   &
$ D^0 \eta$ & &$(2.1\pm1.2) \times 10^{-5}$  \\
$ D^0  \eta^\prime$ & &   $(1.7\pm0.4)\times10^{-4}$ &  
$ D^0 \eta^\prime$ & & $(9.8\pm7.6) \times 10^{-6}$\\
$ D^0K^-$ & $V^*_{us}V_{cb}\,\, (C+T)$ &$(3.7\pm0.6)\times10^{-4}$ &
$ D^0\pi^0$ & $-V^*_{us}V_{cb}\,\, \frac{E}{\sqrt{2}}$&  $(1.0\pm0.3)\times 10^{-6}$ \\
$ D^0\overline{K}^0$ & $V^*_{us}V_{cb}\,\, C$  & $(5.0\pm1.4)\times10^{-5}$&
$ D^+\pi^-$ &$V^*_{us}V_{cb}\,\, E$ & $(2.0\pm0.6)\times 10^{-6}$ \\
$ D^+K^-$ & $V^*_{us}V_{cb} \,\, T$ &$(2.0\pm0.6)\times10^{-4}$ & 
$ D_s^+ K^-$ &$V^*_{us}V_{cb}\,\, (T+E)$ & $(1.8\pm0.3)\times 10^{-4}$ \\
\hline 
    \end{tabular}
    \end{center}
  \end{table*}

Predictions for the masses of excited $B$ mesons,  collected in Table \ref{bmasses}, can be obtained 
if   the splittings  $\Delta_S$ and $\Delta_T$ are the same for charm and beauty, which  is true in the rigorous heavy quark limit;
at $O(1/m_{Q})$ this  corresponds to assuming that the matrix element of the kinetic energy operator is the same for the three doublets.  
 $B_{s0}^*$ and  $B_{s1}^\prime$  turn out to be below the $B K$ and $B^{*} K$ thresholds,
 therefore they  are expected to  be  narrow 
\cite{Colangelo:2003vg,Colangelo:2004vu}.    Preliminary data are available from Tevatron:
$M(B_1)=5734\pm3\pm2$ MeV and 
$M(B_2^*)=5738\pm5\pm1$ MeV (CDF),  and 
$M(B_1)=5720.8\pm2.5\pm5.3$ MeV and 
$M(B_2^*)-M(B_1)\simeq 25.2\pm3.0\pm1.1$ MeV (D0), together with
$M(B_{s2}^*)=5839.1\pm1.4\pm1.5$ MeV (D0) 
\cite{Kravchenko:2006qx}. The $B_2^*-B_1$ mass splitting measured by D0 is compatible with the
prediction. The difference between the predicted and the measured masses is of $O(\Lambda^2({1 \over m_c}-{1 \over m_b}))$, i.e. of the size of the terms  neglected in the calculation. 

The couplings $h^\prime$  and
$f$ in eqs.(\ref{lag-hprimo}-\ref{lmix})  can be obtained from
 the widths of the two members of the     $s_\ell^P={3 \over 2}^+$
doublet,  $D_1$ and $D_2^*$, and those 
of charmed-strange meson transitions:
$D_{s2}^{*+} \to D^{(*)+}K^0$, $D^{(*)0} K^+ $ and $D_{s1}^+ \to D^{*+} K^0, \,D^{*0} K^+$.
We use recent data  from Belle Collaboration \cite{Abe:2003zm}:
$
\Gamma(D_2^{*0})= 45.6 \pm 4.4 \pm 6.5 \pm 1.6
\,\,\, {\rm MeV}$,
$\Gamma(D_1^{0}) = 23.7 \pm 2.7 \pm 0.2 \pm 4.0 \,\,\, {\rm MeV}$,
together with   $h=-0.56$ \cite{Colangelo:1995ph}, a  theoretical estimate
  coinciding  with the value obtained
from the width of $D^*_0$.

 A further constraint is the  Belle
 measurement of the helicity angle distribution in the
decay $D_{s1}(2536) \to D^{*+} K_S^0$,  with the determination of the ratio
$R= \displaystyle{\Gamma_S \over
\Gamma_S+\Gamma_D } $,
$\Gamma_{S,D}$  being the S and D wave
partial widths, respectively \cite{Abe:2005xj}:
 $0.277 \le R \le 0.955$.
Taking into account all the constraints, we get:
\be
h^\prime= 0.45 \pm 0.05 \,\,\,\,\,  f=0.044 \pm 0.044  \,\,{ \rm GeV}\label{results} \,\,.\ee
The coupling constant $f$ is  compatible with zero,  indicating
that the contribution of the  Lagrangian term (\ref{ld1})  is small. Since also
the coupling $g_{1}$ turns out to be small, the two $1^{+}$ states corresponding to the
$s_{\ell}^{P}={1 \over 2}^{+},{3 \over 2}^{+} $
practically coincide with the physical states. We obtain
$\Gamma(D_{s1}(2536))=2.5 \pm 1.6 \,\,\,{\rm MeV}$ and the
widths of excited $B_{(s)}$ mesons   in  Table \ref{larghezze}.
A word of caveat is needed here, since these predictions are obtained only considering
the heavy quark spin-symmetry breaking terms in the effective Lagrangian; corrections due to
spin-symmetric but heavy flavour breaking terms involve additional couplings for which
no  information is currently available, so that they
cannot be reliably bounded.  The estimated widths turn out to be larger than the preliminary measurements \cite{Kravchenko:2006qx}:  a discussion of  this interesting point requires a
confirmation of the experimental data.

\section{A CLASS OF $B_s$ DECAYS BY AN SU(3) ANALYSIS}

Coming to weak interaction processes, it is  again possible to exploit 
the idea of using a symmetry and the experimental data
to make predictions \cite{paper2}.
In this case, the symmetry is $SU(3)_F$  and the predictions 
concern the rates of a class of $B_s$ decay modes, an important topic for  the
 $B_s$ physics programs at the  Tevatron and at  the
 LHC.  
We  consider the  modes induced by the quark
transitions $b \to c \bar u d$ and $b \to c \bar u s$, namely
those collected in Table \ref{tab1}.
 They are governed, in the $SU(3)_F$ limit, by  few independent amplitudes that can be  constrained, both in moduli and  phase differences, from
 $B$ decay data.
Considering transitions  with  a light pseudoscalar meson
belonging to the octet in the final state, there are three different
topologies in decays induced by  $b \to c \bar u d (s)$,  the color allowed topology $T$,
the color suppressed topology $C$ and the $W$-exchange topology
$E$.  The transition in the $SU(3)$ singlet $\eta_0$ involves
another amplitude $D$,  not related to the previous
ones.  The identification of the different amplitudes
can be done observing that
$\overline{B}\rightarrow DP$ decays induced by 
 $b\rightarrow c\overline{u}q$ ($q=d$ or $s$) involve a
weak Hamiltonian transforming as a flavor octet: 
$H_W=V_{cb}V^*_{ud}T^{(8)}_{0\, 1\, -1}+V_{cb}V^*_{us}T^{(8)}_{-1\,\frac{1}{2}\, -\frac{1}{2}}$
(denoting by $T^{(\mu)}_\nu$  the $\nu=(Y,I,I_3)$ component of an
irreducible tensor operator of rank $(\mu)$).
When combined with the initial $\overline{B}$ mesons, which form a
$(3^*)$-representation of SU(3), this leads to $(3^*)$, $(6)$ and
$(15^*)$ representations. These are also the representations
formed by the combination of the final octet light pseudoscalar
meson and triplet D meson. Therefore, the  three reduced amplitudes are $\langle
\phi^{(\mu)}|T^{(8)}|B ^{(3^*)}\rangle$, with $\mu=3^*,6,15^*$  \cite{Zeppenfeld:1980ex}.
Linear combinations of the   reduced amplitudes produce
 the three topological diagrams.

\begin{figure}[htb]
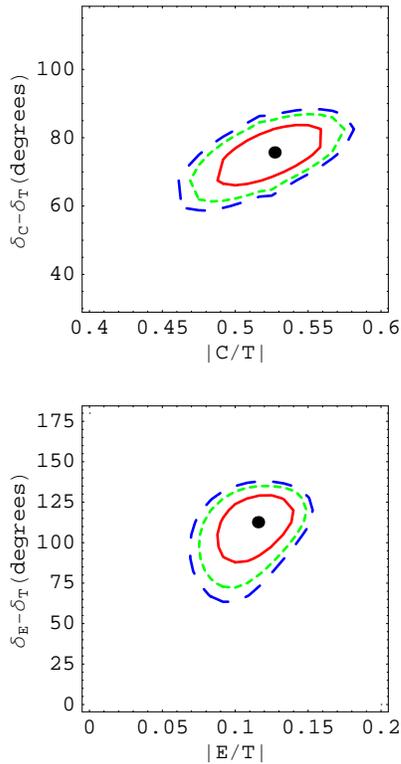

\begin{center}
\includegraphics[width=0.33\textwidth] {CT.eps} 
\includegraphics[width=0.33\textwidth] {ET.eps}
\vskip -0.5cm
\caption{
Ratios of SU(3) amplitudes   obtained from $B$ data in Table \ref{tab1},  at   
 $68\%$ (continuous), $90\%$ (dashed) and $95\%$ CL (long-dashed contour).
The dots are the result of the fit.}
\vskip -0.5cm
\label{fig:BDM}
\end{center}
\end{figure}

The four $\bar B \to D \pi$ and $\bar B \to D_s K$  rates
cannot determine $C$, $T$, $E$ and their phase differences \cite{neub}.
$\bar B \to D_s K$ only
 fixes the modulus of $E$, which is sizeable.
Moreover, the presence of $E$ does not allow to directly relate the amplitudes
 $T$  or   $C$  in $D \pi$ and $D K$.  However, all the
information on $\bar B \to D \pi, D_s K$ and $D K$ can be used
 to tightly determine $T$, $C$ and $E$.  In  fig.\ref{fig:BDM}  the allowed regions in the $C/T$
and $E/T$ planes are depicted. 
 The phase differences between the various amplitudes
are large, showing    
deviation from naive (or generalized) factorization,  providing contraints to QCD-based approaches to non leptonic $B$ decays  
\cite{Beneke:2000ry}
 and suggesting sizeable long-distance effects  \cite{violations}.
Fixing   $|V_{us}/V_{ud}|=0.226\pm0.003$, 
 we obtain $|\frac{C}{T}|=0.53 \pm 0.10$,  $|\frac{E}{T}|=0.115\pm0.020$,
$\delta_C-\delta_T=(76\pm12)^\circ$   and   $\delta_E-\delta_T=(112\pm46)^\circ$.
%
%
%
With the resulting amplitudes  we can predict the rates of
$B_s$ decays in Table \ref{tab1}. The uncertainties in the predicted rates are small, even in
the $W$-exchange induced processes   $\overline{B}_s^0
\to D^+ \pi^-, D^0 \pi^0$. 
For the mode $B_s \to D_s^- \pi^+$,
 the predicted ratio $\displaystyle{ \Gamma (B_s \to D_s^- \pi^+)\over \Gamma (B^0 \to D^- \pi^+)}=1.05\pm0.24$ can be compared to the measurement $\displaystyle{ \Gamma (B_s \to D_s^- \pi^+)\over \Gamma (B^0 \to D^- \pi^+)}=1.32\pm0.18\pm0.38$ 
\cite{CDFBs}.
The decays into  $\eta$ or $\eta'$  
involve the amplitude $D$ corresponding to the transition in the $SU(3)$ singlet $\eta_0$,  and
the $\eta-\eta'$ mixing  angle $\theta$. Fixing  
$\theta=-15.4^0$,  we  obtain $|\frac{D}{T}|=0.41\pm0.11$ and
$\delta_D-\delta_T=-(25 \pm 51)^\circ$, and the rates of corresponding
$\overline{B}_s^0 $ modes.

Among  other $b \to c \bar u d (s)$ induced  modes,  we consider those into
 $D_{(s)}  V$,   $D^*_{(s)}  P$, with  the same $SU(3)$ decomposition
as in Table \ref{tab1}. $B$ decay data are collected in Table \ref{tab3}, including
the recently observed W-exchange mode $\bar B^0 \to D^{*}_sK^{-},$ together with the predictions for
$B_s$.    
Present experimental data for other  modes induced by the same quark
transitions, namely   $\bar B \to D^*_{(s)}ÊV$ decays,
are not precise enough to  constrain the independent $SU(3)$ amplitudes.
  
$SU(3)_F$ breaking  can modify the predictions: its  effects are not
universal  and in
general cannot be reduced to well defined and predictable
patterns. Its parametrization   introduces additional  quantities   that  at present
cannot be sensibly bounded since they are obscured  by  the experimental uncertainties. It will be interesting to
investigate its  role when  other $B_s$ decay rates will be measured and
more precise $B$ branching fractions  will be available. \\

\noindent{\bf Acknowledgments} {\,\,\,We thank the organizers for kind invitation to this interesting
workshop.}
%
\begin{table*}[htb]
\caption{Experimental branching fractions of $\bar B \to D_{(s)}  V, \,\,  D^*_{(s)}  P$ decays and  predictions for $\overline{B}_s^0$.}
    \label{tab3}
    \begin{center}
    \begin{tabular}{l c |  l c}
  \hline 
$B^-, \overline{B}^0$  &   BR (exp) & $\overline{B}_s^0$  & BR (th)\\
\hline
$ D^0\rho^-$ &  $(1.34\pm0.18)\times10^{-2}$&$ D _s^+ \rho^{-}$ &   $(7.2\pm 3.5)\times10^{-3}$ \\
$D^0\rho^0$ &  $(2.9\pm1.1)\times10^{-4}$&$ D^0\overline{K}^{*0}$ & $(9.6\pm2.4)\times10^{-4}$ \\
$ D^+\rho^-$    &  $(7.7\pm1.3)\times10^{-3}$& &  \\
$ D_s K^{*-}$    &  $<  \,9.9 \times10^{-4}$& & \\
 $ D^0 K^{*-}$ &  $(6.1\pm2.3)\times10^{-4}$&$ D^0 \rho^{0}$ &   $(0.28\pm 1.4)\times10^{-4}$ \\
$ D^0 \bar K^{*0}$ &  $(4.8\pm1.2)\times10^{-5}$&$ D^+ \rho^-$ & $(0.57\pm 2.8)\times10^{-4}$ \\
$ D^+ K^{*-}$    & $(3.7\pm1.8)\times10^{-4}$& $ D_s^+K^{*-}$ &  $(4.5\pm3.1)\times10^{-4}$ \\
\hline 
$ D^{*0}\pi^-$ &  $(4.6\pm0.4)\times10^{-3}$&$ D _s^{*+} \pi^{-}$ &   $(3.2\pm 0.2)\times10^{-3}$ \\
$ D^{*0}\pi^0$ &  $(2.7\pm0.5)\times10^{-4}$&$ D _s^{*0} \bar K^{0}$ &   $(4.7\pm 2.2)\times10^{-4}$ \\
$ D^{*+}\pi^-$ &  $(2.76\pm0.21)\times10^{-3}$&&    \\
$ D^{*}_s K^-$ &  $(2.0\pm0.6)\times10^{-5}$& &    \\
$ D^{*0}K^-$ &  $(3.6\pm1.0)\times10^{-4}$&$ D ^{*0} \pi^{0}$ &   $(5.7\pm 1.7)\times10^{-7}$ \\
& &$ D ^{*+} \pi^{-}$ &   $(11.5\pm 3.4)\times10^{-7}$ \\
$ D^{*+}K^-$ &  $(2.0\pm0.5)\times10^{-4}$&$ D _s^{*+} K^{-}$ &   $(1.3\pm 0.2)\times10^{-4}$ \\
\hline
      \end{tabular}
      \end{center}
  \end{table*}
%

\end{document}